\documentclass[twocolumn,epsfig,pre]{revtex4}
\usepackage{graphics}
\usepackage{graphicx}
\usepackage{epstopdf}
\usepackage{amssymb}
\usepackage{amsmath}
\usepackage{hyperref}
\usepackage{latexsym}
\usepackage{color}
\usepackage{multirow}
\usepackage{array}
\usepackage{bm}

\begin{document}

	\renewcommand{\theequation}{\arabic{section}.\arabic{equation}}
	\renewcommand{\thefigure}{\arabic{figure}}
	
	\title{Dynamics of stiff filaments in size-polydisperse hard sphere fluids} 

	\author{Thokchom Premkumar Meitei}
	\author{Lenin S. Shagolsem}
	\email{lenin.shagolsem@nitmanipur.ac.in}
	\affiliation{Department of Physics, National Institute of Technology Manipur, Imphal, India} 
	
	\date{\today}

\begin{abstract}

\noindent The dynamics of a stiff filament (made by connecting beads) embedded in size-polydisperse hard sphere fluid is investigated by means of molecular dynamics simulations with focus on how the degree of size-polydispersity, characterized by polydispersity index ($\delta$), affects the dynamics in this model heterogeneous system. Polydispersity of the fluid as well as strong coupling of rotational and translational motion of the rods are two of the various hurdles in interpreting the experimental results in a complex fluid environment. 
Furthermore, influence of volume fraction, $\phi$, and absolute free volume, $V_{\rm free}$, which changes inherently with $\delta$ on the dynamics are not adequately discussed in the literature. Thus, we investigate the dynamical behaviour of the rods under two different conditions: (i) constant pressure (in which $\phi$ changes with $\delta$), and (ii) constant $\phi$. 
Under constant pressure it is observed that the rotational relaxation time and hence the diffusion constant, $D_R$, varies with rod length as $D_R \sim l^{-\alpha}$, where the value of exponent $\alpha$ increases from $3.0-3.2$ while varying $\delta$ from $0-40\%$. It is observed that the effect of increasing $\phi$ dominates over the effect of increasing $V_{\rm free}$. 
Also there is minimal effect of hydrodynamic interaction among the beads belonging to a rod during rotation, whereas the presence of partial hydrodynamic screening for the motion of the centre of mass is seen. 
On the other hand, for fixed $\phi$ systems, increasing $\delta$ results in increasing $V_{\rm free}$ and thus enhances tracer diffusion, an opposite trend compared to the system under constant pressure.

\end{abstract}

\maketitle
%%\tableofcontents

%%%%%%%%%%%%%%%%%%%%%%%%%%%%%%%%%%%%%%%%%%%%%%%%%%%%%%%%%%%%%%%%%%%%%
\section{Introduction}
\label{sec: intro}
Stiff filaments or nano-rods can mimic Tobacco Mosaic Virus and segments of DNA/RNA,\cite{supakitthanakorn2022tobacco,tirado1984comparison,cush2004rotational} and are a promising drug delivery cargo as they posses enhanced diffusivity compared to the spherical counterparts \cite{yu2016rotation,bao2020enhanced,wang2018diffusion,wang2017molecular}. Modification of nano-rods with tunable/enhanced mechanical \cite{koerner2009zno,mccook2006epoxy,shen2018mechanical} and electrical \cite{brioude2005optical,jain2006plasmon} properties affects their dynamical behavior in complex heterogeneous media such as polymer composites and biological fluids\cite{bao2021experimental}. 
Additionally, biological fluids like intracellular environment consist of different constituents\cite{jeon2016protein,sokolov2012models,shagolsem2022energy} with varying shape, size, and interaction energy. 
Despite many significant advances in the field, the detailed mechanism of the rod's motion in biological fluids is still elusive because of factors like polydispersity, aggregration of rods, etc.~\cite{wang2021diffusion} \

In this study, a complex heterogeneous medium is modeled by size-polydisperse hard-sphere (HS) fluid system and explore the dynamics of rods embedded in such fluid medium for varying degree of size-polydispersity. 
The dynamics of rod exhibits different modes, namely, translational motion (along/normal to the rod axis) and rotational motion. Their corresponding diffusion coefficients depend on the bulk viscosity of the fluid medium. However, recent studies suggest that  nano-rods couple to a fraction of the bulk viscosity only,\cite{alam2014translational,choi2015fast,li2018translational,karatrantos2019nanorod} and the relative size of the rods and the fluid particles  play a pivotal role in the determination of the interaction between the rods and the fluid medium. 
Dependending on the rod size/length ($l$) w.r.t.~the correlation length ($\xi$) of the background fluid medium it may feel bulk or local viscosity, e.g., for rods in polymer melts, when $l \gg \xi$ it feels bulk viscosity, while for $l \approx \xi$ it experiences local viscosity whose value depends on the nature of interaction between the rod and the fluid particles \cite{choi2015fast,brochard2000viscosity}.  
Generally, an attractive interaction between rod and fluid results in the increase in the friction offered by the fluid medium on the rods, thereby resulting in negative deviation from Stoke-Einstein (SE) equation. And a repulsive interation generally results in the positive deviation from SE equation. This breakdown of SE relation is also expected in the size-polydisperse fluids with inherent spatial heterogenity of the bulk. 
On the other hand, the dynamics of such filaments are also affected by their shape and mechanical poperties. For example, it is observed that even a small bending flexibility of single walled carbon nanotubes (SWCNTs) confined in an agarose gel network strongly enhances their motion: the rotational diffusion constant is proportional to the filament bending compliance, while the mobility can be controlled by tuning the flexibility.\cite{fakhri2010brownian}
In this context, molecular dynamics (MD) simulations provide a convinient way to decouple rotational and translational modes of motion and we study each mode of motion independently. 

 To describe the dynamics of flexible and stiff polymer chains in a crowded environement, de Gennes,\cite{de1971reptation} Doi and Edwards\cite{doi1979dynamics,edwards1968statistical} introduced the reptation model and  its consequences were validated through computer simulations and experimentals by direct imaging of fluorescently labeled DNA \cite{perkins1994direct} and actin \cite{kas1994direct}. The rotational diffusion constant of a rod of length $l$ and thickness or diameter $b$ depends on the concentration of rods $c$, which is defined as the number of rods per unit volume. In the dilute limit ($c<1/l^3$), where the interaction between the rods is limited, the rotational dynamics of the rods is mainly determined by the hydrodynamic friction exterted by the fluid particles and $D_R= k_BT\ln(2l/b)/(3\pi\eta_sl^3) $\cite{de1971reptation,kirkwood1951visco,flory1956phase,riseman1950intrinsic} with $\eta_s$ the fluid viscosity. However, as the rod concentration increases, the interaction among the rods (steric repulsion) becomes too important a factor to be neglected. However, the system is isotropic upto a concentration $c_1 \approx 1/bl^2$, beyond which isotropic-nematic transition occurs\cite{doi1975rotational}. In the concentrated regime ($c>1/l^3$), a monomerically thin rod can easily translate along its axis, while the motion along the direction normal to the rod axis experiences steric hindrances and due to which it is confined to move in an imaginary cylindrical space, along whose axis the rod reptates. However, the dynamics of thermal motion of stiff rods in a network is yet to be fully understood.\cite{fakhri2010brownian} Morever, there are contrasting theoretical predictions pertaining to the role of bending stiffness of such rods on the rotational diffusion.\cite{doi1975rotational,hofling2008entangled,odijk1983statistics,ramanathan2007simulations,sato1991dynamics}  While Doi\cite{doi1975rotational} predicted that flexibility has no effect on the rotational diffusion of stiff rods, Odijk\cite{odijk1983statistics} advocated enhancement of rotational diffusion on inclusion of bending flexibility in contrast to the views of Sato\cite{sato1991dynamics}. However, in the direct visualisation of SWNTs using near-infrared video microscopy\cite{fakhri2010brownian}, the rotational diffusion was profoundly enhanced because of bending flexibility in agreement with Odijk. In addition to these, experimental findings by means of birefringence and
dichroism\cite{tracy1992dynamics,wijmenga1986rotational,zero1982rotational}  are conflicting, with the major reason being polydispersity, aggregration of rods and strong correlation between rotation and translational motions which complicate the interpretation of results.\cite{fakhri2010brownian}. \

In this study, we consider monomerically thin stiff nanorods embedded in size-polydisperse fluids in the limit of low rod concentration, i.e., dilute regime. The degree of size disparity of the constituent HS fluid particles is characterised by the polydispersity index $\delta$ (defined as the ratio of the standard deviation to the mean of the distribution). We systematically explore the effect of varying degree of size-polydispersity of the fluid particles on the translational and rotational dynamics of the rods of different lengths. The extent of hydrodynamic screening between the monomers of the same rod is also investigated, which is necessary as there is presently no theory \cite{wang2021diffusion} (to the best of our knowledge) describing the friction co-efficient of motion perpendicular to rod axis as a function of fluid viscosity. \

The rest of the manuscript is organized as follows. In section~\ref{sec: model-description} we present the model and simulation set up; the results of the investigation of rotational and translational motions of the rods are discussed in section~\ref{sec: rotation} and \ref{sec: Translation} respectively, and finally we conclude in section~\ref{sec: conclusion}. 

%%%%%%%%%%%%%%%%%%%%%%%%%%%%%%%%%%
%%%%%%%%%%%%%%%%%%%%%%%%%%%%%%%%%%
\begin{figure}[!h]
	\includegraphics[width=0.95\linewidth]{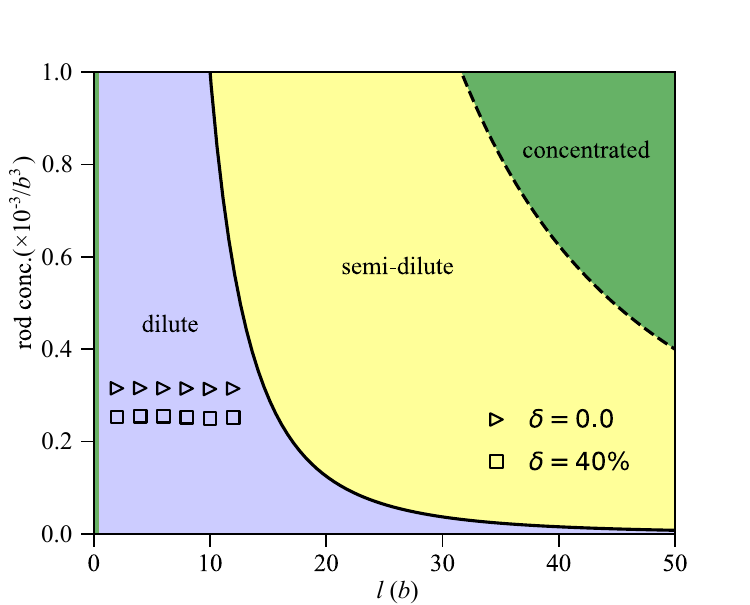}
	\caption{Rod concentration of the simulated system for different rod length. Regions: Blue -- dilute concentration limit($c < 1/l^3$), Yellow -- semi-dilute ($1/l^3< c< 1/(bl^2)$), and Green -- concentrated limit ($c > 1/(bl^2)$). Rod concentrations considered in this study are indicated for $\delta = 0,~40\%$.}
	\label{fig: rod_conc}
\end{figure}

\section{Model and simulation details}
\label{sec: model-description}
Dynamics of stiff rods embedded in size-polydisperse hard spheres fluids are modeled via coarse-grained molecular dynamics (MD) simulations. The concentration of rods, i.e.~the number of rods per unit volume, considered is in the dilute limit as shown in the figure \ref{fig: rod_conc}. In particular, we consider a size-polydisperse hard sphere fluid system consiting of $N=5000$ particles in which the rods are embedded. The sizes of the fluid particles are assigned through random sampling from Gaussian distribution, 
\begin{equation}
	\label{eqn:normal-dist}
	P(\sigma)=\frac{1}{\beta\sqrt{2\pi}}\exp\left[-\frac{1}{2}\left(\frac{\sigma-\bar{\sigma}}{\beta}\right)^2\right]~,
\end{equation}
with $\beta$ the standard deviation and $\bar{\sigma}=2$ the mean particle size. The extent of size disparity is characterized through polydispersity index, $\delta = \beta/\bar{\sigma}$ and we vary $\delta$ upto $40\%$. \
All the pair-wise interaction in the system is modeled through Lennard-Jones (LJ) potential, 
\begin{equation}
U^{\rm{LJ}}_{\tiny{_{ ij}}}(r) = 4\epsilon_{ij}\left[\left(\frac{\sigma_{ij}}{r_{ij}}\right)^{12}-\left(\frac{\sigma_{ij}}{r_{ij}}\right)^{6}\right]~,
\label{eqn: LJ-potential}
\end{equation}
where $\sigma_{ij}$ is the arithmetic mean of diameters of $i^{\rm th}$ and $j^{\rm th}$ particles, $r_{ij}$ and $\epsilon_{ij}$ are the pair-wise separation and interaction energy parameter, respectively, between the particles. The LJ potential is cut at the potential minima, i.e., $r_{\rm cut} = r_{ij} = 1.12\sigma_{ij}$ and shifted to zero so as to represent only the steric repulsion between the particles. 
The energy parameter $\epsilon_{ij}$ is the same for every pair of particles, and it is set to $\epsilon_{ij}=\epsilon=1$, also the mass of each coarse-grained beads (irrespective of size) is set to unity. 
\begin{figure}
\includegraphics[width=0.95\linewidth]{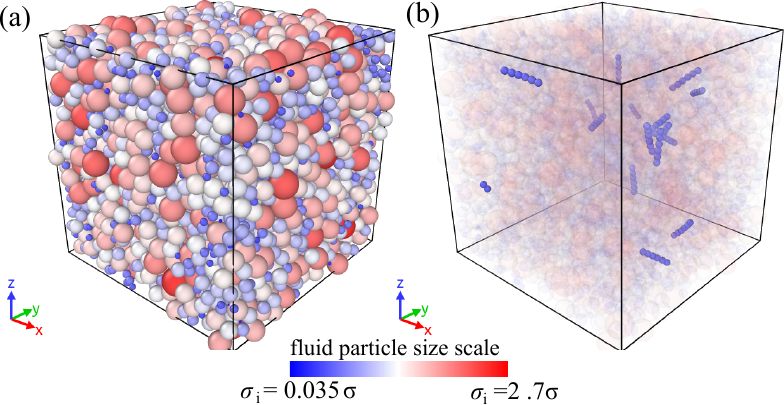}
\caption{Typical simulation snapshot of 15 rods of length $l/b=6$ each embedded in a size polydisperse fluid medium of $\delta = 40\%$. (a) Size polydisperse fluid medium with particle size ranging from $\sigma_i = 0.035\sigma-2.7\sigma$. (b) The fluid particles are made transparent to show the rods. The diameter of the monomers of each rod is set to $b = 0.5\sigma$.}
\label{fig: snapshot}
\end{figure}
The rods are made of coarse-grained beads (each of diameter $b=\sigma=\bar{\sigma}/2$ and mass = 1) which are connected via the finite extensible nonlinear elastic (FENE) bond potential, 
\begin{equation}
U_{\rm{FENE}}=-0.5K_{\rm{F}}R_0^2{\ln}\left[1-\left(\frac{r}{R_0}\right)^2\right]
\end{equation}
with $K_{\rm F}$ the spring constant, $R_0$ the maximum bond extension. The angle between two successive bonds are modelled by harmonic potential,  
\begin{equation}
\label{eqn: harmonic angle}
U_{\rm{H}}=K_{\rm{H}}(\theta-\theta_0)^2,
\end{equation}
where $K_{\rm H}$ is the spring constant (anlogous to bending stiffness) and the factor 1/2 has been absorbed into it and it is set to 300 (in units of $k_BT$), $\theta$ being the angle between two successive bonds and $\theta_0$ its equillibrium value which is set to $\pi$ rad. In this study, the length of rods varies in the range $l/b=2-12$, and the number of rods is taken to be 15. The rod-rod and rod-fluid interactions are modeled by hard sphere potential (eqn.~\ref{eqn: LJ-potential}) and the agglomeration of rods is absent in the considered dilute limit of rod concentration. All the physical quantities are expressed in LJ reduced units, where length in the units of $\sigma$, temperature $T$ in $\epsilon/k_B$, pressure in $\epsilon/\sigma^3$, and time in $\tau_{_{LJ}}=\sqrt{m\sigma^2}/\epsilon$.\cite{allen2017computer} \\

The equation of motion of the $i^{\rm th}$ particle is given by the Langevin equation 
\begin{equation}
\label{eqn: Lan}
m_i\frac{\text{d}^2{\textbf{r}_i}}{\text{d}t^2}+\zeta\frac{\text{d}{\textbf{r}_i}}{\text{d}t}=-\frac{\partial U}{\partial{\textbf{r}_i}}+{\textbf{f}_i}(t),
\end{equation}
with ${\textbf{r}_i}$ the position of $i^{th}$ particle, $\zeta$ the friction coefficient, \textit{U} the net pair potential, and ${\textbf{f}_i}(t)$ a random external force which satisfies the relations: $\langle {\textbf{f}_i}(t)\rangle=0$ and $\left\langle {\textbf{f}_i}^\alpha(t){\textbf{f}_j}^\beta(t)\right\rangle=2\zeta m_ik_BT\delta_{ij}\delta_{\alpha\beta}\delta(t-t')$, where $\alpha$ and $\beta$ are cartesian components. The friction coefficeint $\zeta=1/\tau_d$, $\tau_d$ being the characteristic viscous damping time which is set to 50 in these simulations. Integration of the equation of motion were performed using velocity-Verlet scheme with time step $\delta t=0.005$.\cite{allen2004introduction}.
%We perform MD simulations in NPT ensemble using Nos\'{e}-Hoover thermostat and barostat \cite{todd2017nonequilibrium} to maintain constant temperature and pressure, respectively. 
Initially, the systems are prepared under constant pressure ($P=1$) and temperature ($T=1$) using Nos\'{e}-Hoover thermostat and barostat \cite{todd2017nonequilibrium}, and after well equilibration we switch the systems to constant NVT to investigarte various dynamic properties.  
In the considered range of $\delta~(0-40\%)$, the volume fraction, $\phi$, of the system roughly varies in the range $0.44-0.53$.  
A typical simulation snapshot of the system at $\delta=40\%$ is shown in figure \ref{fig: snapshot}. Note that when polydispersity index changes new particle sizes are introduced in the system leading to a different volume fraction for different value of $\delta$. 
%%Here, changing $\delta$ inherently introduce change in volume fraction (under NPT condition). 
In order to decouple this effect, we prepare samples at fixed volume fraction irrespective of the value of $\delta$. In this case, to maintain a fixed value of $\phi$, at different values of $\delta$, the systems' volume are adjusted accordingly and relaxed under NVT condition. 
Finally, in order to obtain statistically accurate results, we considered at least 15 replicas for each value of $\delta$ and the results reported here are averaged values. 
%\newpage

%%%%%%%%%%%%%%%%%%%%%%%%%%%%%
%%%%%%%%%%%%%%%%%%%%%%%%%%%%%
\section{Rotational Dynamics} \label{sec: rotation}

\begin{figure}
	\includegraphics[width=0.95\linewidth]{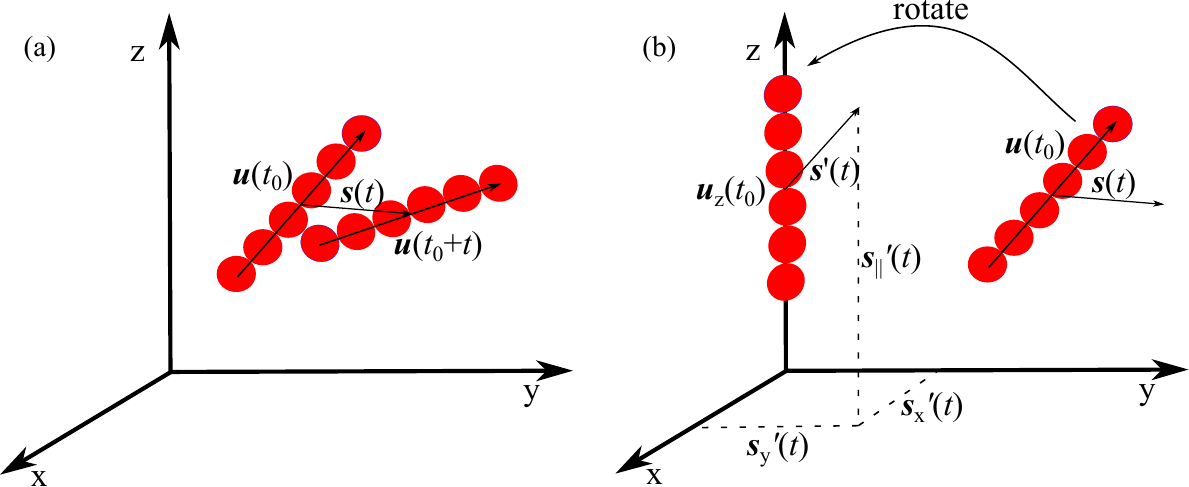}
	\caption{\small(a) Illustration of orientational unit vectors $\hat{\textbf{u}}(t_0)$ and $\hat{\textbf{u}}(t_0+t)$ of rod at time $t_0$ and $t_0+t$ respectively. $\textbf{s}(t)$ represents the displacement of the centre of mass of the rods during the time interval $t$ in the lab frame of reference. (b) The orientation vector $\hat{\textbf{u}}(t_0)$ is rotated such that its axis is aligned along the z-axis, the corresponding transformation is performed on $\textbf{s}(t)$ to obtain $\textbf{s}'(t)$ which is further resolved into components $\textbf{s}'_{||}(t)$, $\textbf{s}'_{x}(t)$ and $\textbf{s}'_{y}(t)$ in the rotated system.}
	\label{fig: rotation}
\end{figure} 

In order to investigate the rotational dynamics of the rods, we follow  the instantaneous orientation vector, $\bold{u}$ (i.e., end-to-end vector) of each rod. As shown in figure \ref{fig: rotation}(a), the rod orientation at times $t_0$ and $t_0+t$ are defined as $\textbf{u}(t_0)$ and $\textbf{u}(t_0+t)$, respectively. 
The rotation of the rods are quantified, in the particle frame of reference, by calculating the Mean Square Angular Displacement (MSAD)\cite{wang2021diffusion} defined as
\begin{equation}
\label{eqn:Ct}
\tilde{C}(t)=\left\langle(\hat{\textbf{u}}(t_0)-\hat{\textbf{u}}(t_0+t))^2\right\rangle.
\end{equation}
 $\tilde{C}(t)$ is related to the time autocorelation function of the rod axis orientation vector by the relation $\tilde{C}(t)=2-2C(t)$,  where $C(t)=\left\langle \hat{\textbf{u}}(t_0)\cdot\hat{\textbf{u}}(t_0+t)\right\rangle$. In the early times (i.e., $t<\tau_{_R}$ with {$\tau_{_R}$ being 1/$D_R$}), the MSAD expression becomes 
\begin{equation}
\label{eqn:rotation_ballistic}
\tilde{C}(t)=2-2\exp(-(D_R/\tau_{_R})t^2)
\end{equation}
and is said to be ballistic. Here, $D_R$ is defined as the rotational diffusion constant. On the other hand, at late times (i.e., $t \gg \tau_{_R}$) the motion is diffusive with the expression of MSAD as 
\begin{equation}
\tilde{C}(t)=2-2\exp(-2D_Rt).
\label{eqn: rotation_diffusive}
\end{equation}
The ballistic and the diffusive regime is interpolated by the approximated expression as sugggested by Wilkinson and Pumir\cite{wilkinson2011spherical},
\begin{equation}
\label{eqn: Wilkinson}
\tilde{C}(t)=2-2\exp\left[ -\frac{2D_Rt^2}{\sqrt{4(1/\gamma)^2+t^2}}\right] ,
\end{equation}
where $\gamma$ is damping factor and [$\gamma]=[t]^{-1}$.
The rotational diffusion constant is extracted from the simulation data by fitting equation \ref{eqn: rotation_diffusive}. 
For the rotation of a thin rod of length $l$ in a fluid of viscosity $\eta_s$, in the absence of hydrodynamic interaction between monomers of the same rod, the expression of friction coefficient and diffusion constant reads~\cite{doi1988theory}:
\begin{equation}
	\begin{aligned}
		\zeta_r&\approx \frac{\pi \eta_sl^3}{4},~{\rm and}~D_R=\frac{4k_BT}{\pi\eta_sl^3},
	\end{aligned}
	\label{eqn: Dr_wh}
\end{equation}
respectively. 
If the hydrodynamics interaction among the beads are taken into account, the motion of a monomer is affected by that of the other monomers of the same rod and in that case, a more accurate relation~\cite{doi1988theory} is given by 
\begin{equation}
	\begin{aligned}
		\zeta_r=\frac{\pi \eta_sl^3}{3\ln{(2l/b)}},~{\rm and}~D_R=\frac{3K_BT\ln{(2l/b)}}{\pi \eta_sl^3}. 
	\end{aligned}
	\label{eqn: Dr_wh_1}
\end{equation} \\

\begin{figure}[!h]
\includegraphics[width=0.9\linewidth]{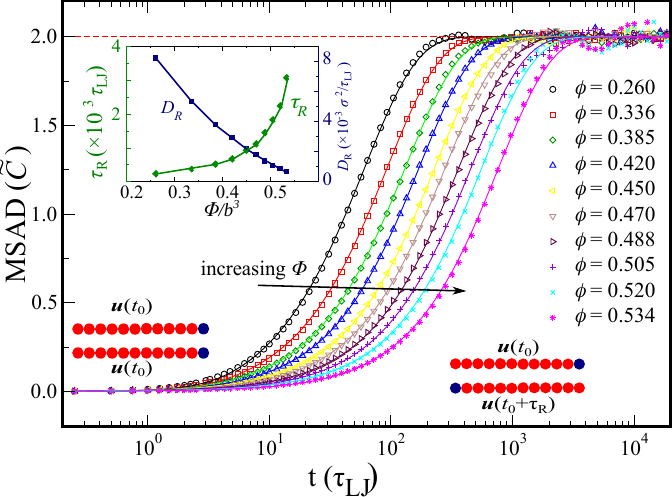}
\caption{The Mean square angular displacement(MSAD) of the the rods of length $l=12b$ embedded in a medium of polydispersity index $\delta=10\%$. The relative orientations of a rod at time $t_0$ and after the elapse of relaxation time  $\tau_{_R}$ are represented by  unit vectors $\hat{\textbf{u}}(t_0)$ and $\hat{\textbf{u}}(t_0+\tau_{_R})$ as shown by the rods made up of beads. Here, the blue bead represents the head of the rod. The solid lines represent best fit to the MD simulation data for various volume fractions ($\phi=0.26 - 0.53$) using equation~\ref{eqn: rotation_diffusive}. (Inset) Variation of $D_R$ and $\tau_R$ with $\delta$. These data are extracted from the MSAD curved using no-linear curve fitting using equation \ref{eqn: rotation_diffusive}.}
\label{fig:MSAD_l12}
\end{figure}

In figure~\ref{fig:MSAD_l12} we display the MSAD of the rods of length $l=12b$ embedded in the size-polydisperse fluid with $\delta=10\%$ shown for different values of system's volume fraction, $\phi=0.26-0.53$. 
The  rods rotate due to the collision with the surrounding fluid particles across the length of the rod and thereby generates torque. However, this random collisions also tend to slow down the rotation. In early times, the rods do not get enough time to undergo multiple collisions and hence they are able to maintain their original orientation ($\tilde{C}=0$). As a result, all the data points collapse to a single line in the ballistic region ($t \ll \tau_{_R}$) of each MSAD curve, suggesting little(negligble) effect of the density of the fluid on the ratation of rods in this region.  The effect of the fluid density is observed once the MSAD completes the ballistic regime but before reaching the late time diffusion limit. In the late time diffusion limit, the rods have undergone multiple collisions and the orientational time auto-correlation saturates at $\tilde{C}=2$. And once again, all the data point collapse into single straight line corresponding to $\tilde{C}=2$.

\begin{figure}[h]
\includegraphics[width=0.95\linewidth]{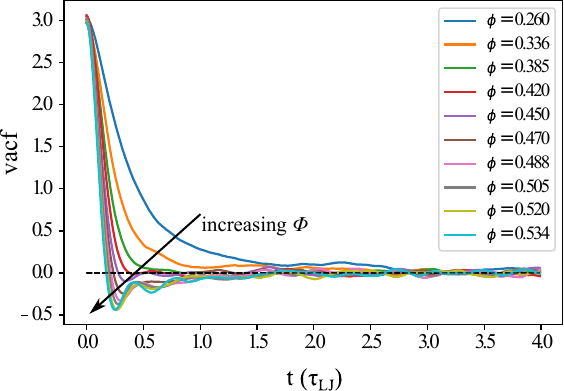}
\caption{The velocity auto corelation function (VACF) of the size-polydisperse fluid particles (with $\delta=10\%$) shown for various volume fractions ($\phi$). The VACF decays faster at larger value of $\phi$  indicating a relatively strong interparticle interaction. Also the area of the curve below zero (see horizontal dashed line at zero for reference) increases with increasing $\phi$.}
\label{fig:vacf_solvent}
\end{figure}
As  density of the fluid medium is increased the rods experience more drag from the surrounding fluid  particles, thereby increasing the relaxation time. This can be inferred from figure \ref{fig:MSAD_l12} where the time taken by the rod axis to rotate $\pi$ rad from its initial position increases as the volume fraction ($\phi$) of the fluid increases from $\phi=0.26$ to $\phi=0.53$. This trend is evident from the figure \ref{fig:MSAD_l12}-inset where $\tau_R $ increases with $\phi$ while the opposite trend is seen for $D_R$.
This observation is also further supported by the velocity autocorelation function (VACF) of the fluid particles, see figure \ref{fig:vacf_solvent}. Since drag force is inversely proportional to the time integration of the VACF the observed increase of negative area (or faster decay of VACF) with increasing $\phi$ indicates a larger drag force offered by the fluid medium with increasing density\cite{seyler2023molecular}. The negative dip in the VACF indicates the reversal of the partcle velocity from its original path indicating backscattering of fluid particles possibly due to the surrouding "cage".
\begin{figure}
\includegraphics[width=0.98\linewidth]{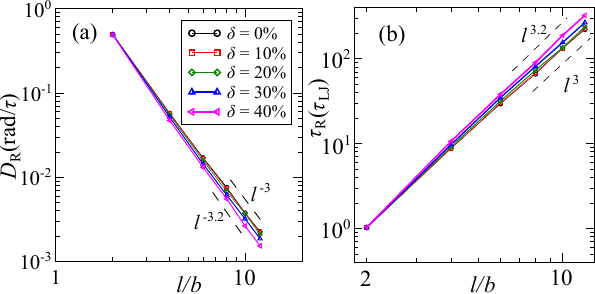}
\caption{(a) The rotational diffusion constant($D_R$) of rods with length $l=2b$ to $l=12b$ in  size polydisperse fluids with polydispersity index ranging from $\delta=0$ to $\delta=40 \%$.(b) The rotational relaxation time of the corresponding systems are shown. The solid lines indicate best power law fits and the black dash lines are guide to the eyes. Each data point corresponds to average of at least 15 realizations of each system. The error is less than the size of the symbols.}
\label{fig:Tr_Dr}
\end{figure}

\begin{figure}
\includegraphics[width = 0.9\linewidth]{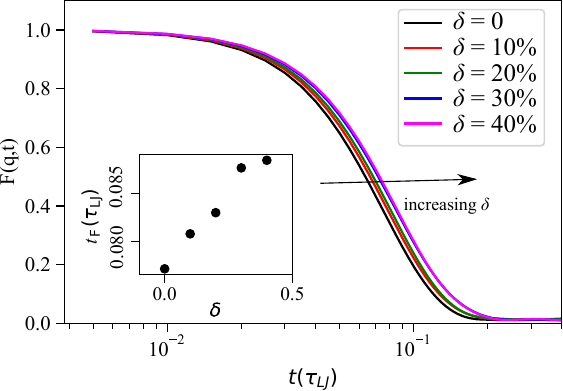}
\caption{Self intermediate scattering function, $F(q,t)$, of the medium with polydispersity indices, $\delta=0.0-0.4$ for $q$ vector corresponding to the first primary peak of the radial distribution function. We observe single relaxation process, as expected, for fluid systems. The data are fitted with streched exponential function $F(q,t) = B\exp[-(t/\tau_F)^\beta]$, where $\tau_{_F}$ is the relaxation time corresponding to translation of a diameter's length of the fluid particle. Inset figure: $\tau_{_F}$ is is plotted as a function of $\delta$.} 
\label{fig:sisf}
\end{figure}

\begin{figure}
\includegraphics[width = 0.5\textwidth]{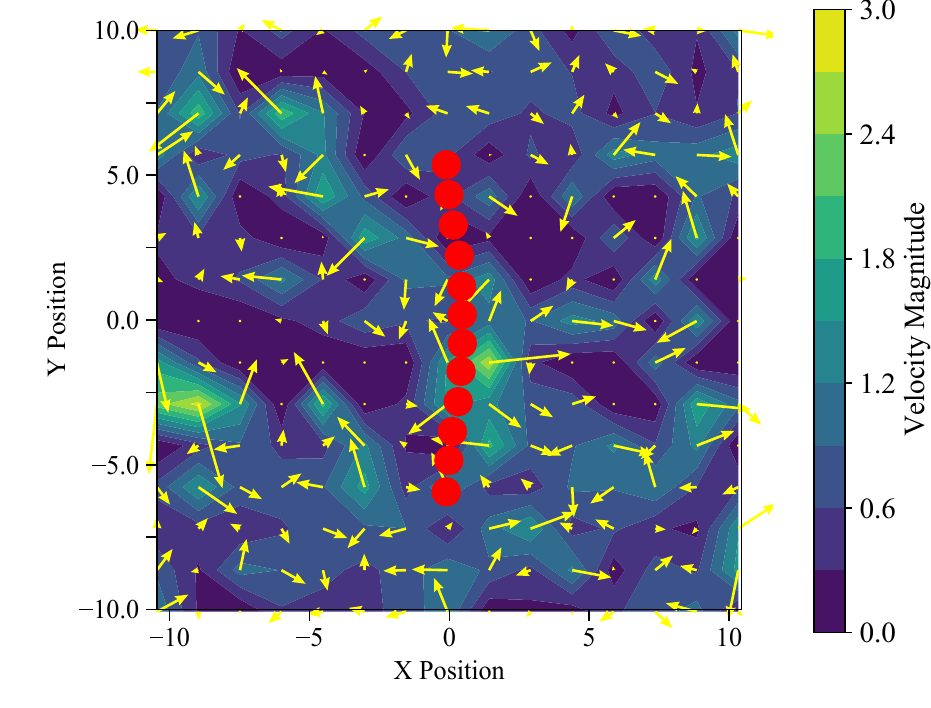}
\caption{Flow field of the particles around a chosen rod(red spheres) of length $l= 12b$  in the rod frame of reference and projected on the plane passing through the centre of mass of the rod(origin) and containing the velocity vector and the orientation vector of the rod. The time averaging  of the particle velocities is done over a time period comparable to the relaxaton time of the rod. And the whole system is already rotated so that the rod aligns along y-axis all the time. The velocity vectors (arrows) of the fluid particles are random in magnitude and direction and a flow field is not observed in the vicinity of the rod.}
\label{fig: flow field}
\end{figure}

The rotational diffusion coefficient($D_R$) and the corresponding rotational relaxation time($\tau_{_R}$) of the rods as a function  their length is plotted in figure \ref{fig:Tr_Dr}. We observe that $D_R\propto l^{-\alpha}$ and $\tau_{_R}\propto l^{\alpha}$ with $\alpha=3$ for the reference monodisperse system($\delta = 0$). This observation is in agreement with equation (\ref{eqn: Dr_wh}), which assumes that there is no hydrodynamic interaction among the monomers of a same rod. 
It is also observed that the magnitude of exponent $\alpha$ increases from $3 - 3.2$ (approximately) as $\delta$ increases from $0-40\%$ indicating a slowing down of the rotational motion for more polydisperse system. 
This observation may be understood as follows. The systems prepared under constant pressure have volume different for different values of $\delta$, where the fluid volume fraction ($\phi$) increases with increasing $\delta$ and hence the relaxation time of the fluid, $\tau_{_F}$ increases, see inset of figure~\ref{fig:sisf}. This effect is amplified, i.e.~slower rotational dynamics or smaller value of $D_R$, for longer rod length as can be seen in figure \ref{fig:Tr_Dr}. \
Also note that the rotational relaxation time is significantly larger than that of the relaxation time of the fluid medium $(\tau_{_F})$, i.e., we find $\tau_{_R}/\tau_{_F} \approx 10^2-10^4$ (for $l=2\sigma-12\sigma$). Thus, a correlated velocity field (of the fluid particles) in the immediate vicinity of the rod is absent, see random velocity distribution of the fluid particles around the rod in figure~\ref{fig: flow field}. \\
In the present study, the rod size is comparable or larger than the mean size of the fluid particles (by a few factors), while in a recent study of rod-shaped polymer in an unenetangled polymer network,\cite{wang2021diffusion} a crossover from $D_R\sim l^{-4}$ to $l^{-3}$ is observed as the size of the rods increases and becomes comparable to that of the polymers in the network.  

%%%%%%%%%%%%%%%%%%%%%%%%%%%%%
%%%%%%%%%%%%%%%%%%%%%%%%%%%%%

\section{Transational Dynamics} \label{sec: Translation}

The overall translational motion is quantified by mean-square displacement (MSD), $\langle \Delta \textbf{r}^2(t)\rangle = \langle [\textbf{r}_{\rm com}(t)-\textbf{r}_{\rm com}(0)]^2\rangle$,  of the centre of mass of each rod. The translational diffusion coefficient ($D_{\rm COM}$) is calculated using the relation $D_{\rm COM}=\langle \Delta \textbf{r}^2(t)\rangle/6t$ in the limit of large $t$. The translational diffusion consists of two modes, one  parallel  and the another perpendicular to the axis of the rod. The parallel component represents the motion of the centre of mass of the rod parallel to its axis while the perpendicular component the represents the motion perpendicular to the axis of the rod. As shown in figure \ref{fig: rotation}, the vector $\textbf{s}(t)$ represents the displacement of the centre of mass of the rod in lab frame. A rotaional transformation is applied on the orientation vector of the rod such that it is oriented along the z-axis, the corresponding transformation is applied to $\textbf{s}(t)$ to obtain $\textbf{s}'(t)$ as shown in fig. \ref{fig: rotation}(b). $\textbf{s}'(t)$ is again resolved into three components: $\textbf{s}_\parallel(t)$ which is  parallel to the rod axis, and $\textbf{s}'_x(t)$ and $\textbf{s}'_y(t)$ which are perpendicular to the rod axis. The diffision coefficient of the parallel component of the motion $D_\parallel$ is obtained from the relation $\langle \textbf{s}'^{2}_\parallel(t)\rangle/2t$, while $D_\perp$ is calculated from $\langle \textbf{s}'^{2}_{x}(t)+\textbf{s}'^{2}_{y}(t)\rangle/4t$ in the long time limit. Thus, the anisotropic constant $k=D_\parallel/D_\perp=2$ in the continuum limit. 

As in the case of rotation, we arrive at a relation for friction coefficient experienced by the rod through simple arguments. Suppose, the friction coefficient experienced by a monomer moving in a medium of viscosity, $\eta$, is $\zeta_0\approx\eta b$. If the rod of length $l$ has $l/b$ beads and no hydrodynamics interaction is present between the beads and the friction experienced by the beads are independent, then the friction coefficient experienced by the rod is $\zeta=(l\zeta_0/b) \approx \eta l$. This friction coefficient has two contributions, i.e., $\zeta_\parallel$ and $\zeta_\perp$ for motion along and perpendicular to the rod axis, respectively, and in an isotropic medium $\zeta_\perp=2\zeta_\parallel$. Thus, in the absence of hydrodynamic interactions or full screening, the diffusion constant $D\propto l^{-1}$. \

If there is hydrodyanamic interaction between the beads of a rod, according to Kirkwood theory\cite{kirkwood1951visco,riseman1950intrinsic}, the friction coefficients for parallel and perpendicular components are, respectively, 
\begin{equation}
\begin{aligned}
\zeta_\parallel &= \frac{2\pi\eta l}{\ln{(2l/b)}}, \quad D_\parallel = \frac{k_BT}{2\pi\eta l}\ln{(2l/b)},
\end{aligned}
\label{eqn: Dcom parallel}
\end{equation}
and
\begin{equation}
\begin{aligned}
\zeta_\perp &= \frac{4\pi\eta l}{\ln{(2l/b)}}, \quad D_\perp = \frac{k_BT}{4\pi\eta l}\ln{(2l/b)}.
\end{aligned}
\label{eqn: Dcom perp}
\end{equation}
And for the centre-of-mass, we have
\begin{equation}
\begin{aligned}
\zeta &= \frac{6\pi\eta l}{\ln{(2l/b)}}, \quad D_{\rm COM} = \frac{k_BT}{6\pi\eta l}\ln{(2l/b)}.
\end{aligned}
\label{eqn: Dcom}
\end{equation}

In the dilute limit ($c<1/l^3$), considering the end effects, the tranlational diffusion constant of the centre of mass of a rigid rod is given by\cite{broersma1981viscous,broersma1960viscous}
\begin{equation}
\label{eqn: Dt}
D_{\rm COM} = \frac{k_BT}{3\pi\eta l} \left( \ln p +\nu \right)
\end{equation}
where $p=l/b$ and $\nu$ is end-effect correction, and a polynomial approximation,\cite{de1984dimensions} 
\begin{equation}
\label{eqn: nu}
\nu = 0.312 +0.565p^{-1}-0.100p^{-2}.
\end{equation}

\begin{figure*}[htbp]
\includegraphics[width=0.9\linewidth]{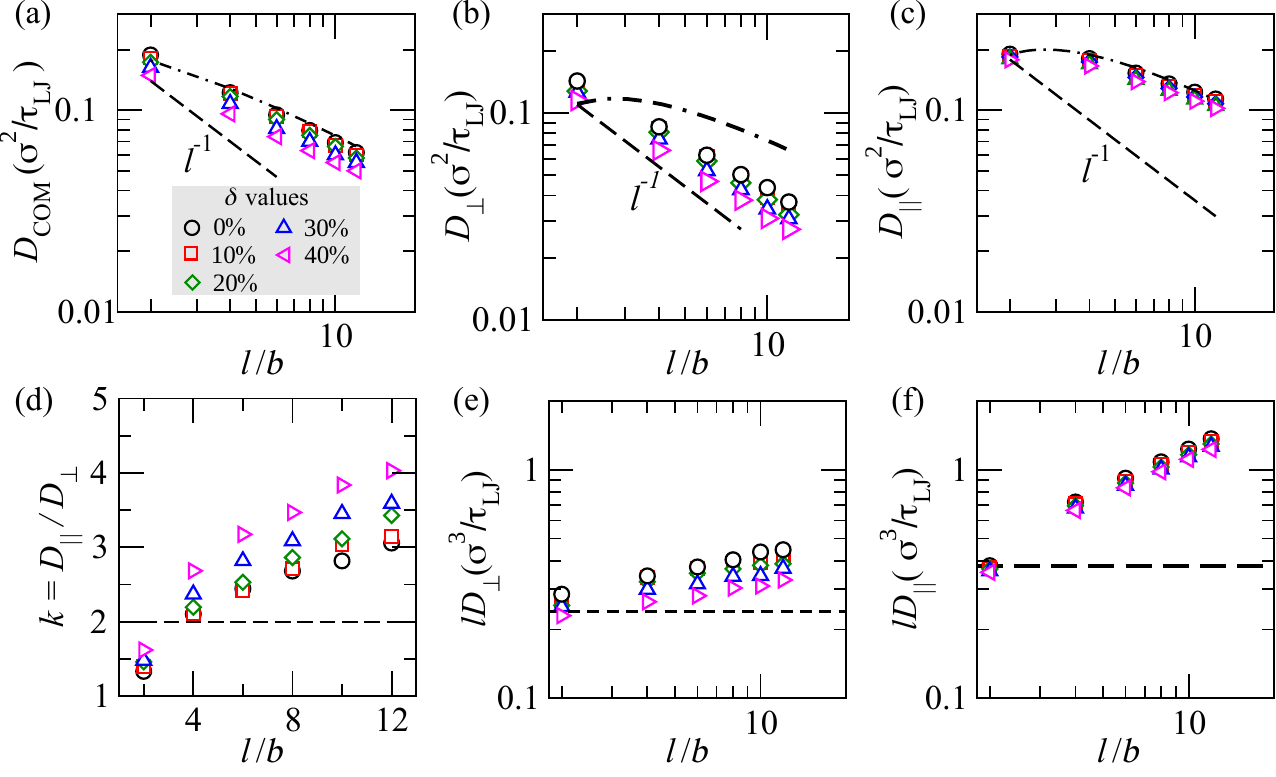}
\caption{The translational diffusion constant as a function of rod length $l$ for (a) centre of mass of the rod, $D_{\rm{COM}}$, (b) motion perpendicular to rod length, $D_\perp$, (c) motion along the long axis of the rod, $D_\parallel$. In figure (a)-(c), the dashed-dot lines represent theoretical prediction for the reference monodisperse fluid with hydrodynamic interaction among the beads of a rod (see eq.~\ref{eqn: Dcom parallel}, eq.~\ref{eqn: Dcom perp}, and eq.~\ref{eqn: Dt}), while dashed line ($\sim l^{-1}$) represents absence of hydrodynamics interaction. 
(d) The anisotropic constant, $\kappa$ vs $l$ at different values of  $\delta$ indicated in the figure. The horizontal dashed line at $k=2$ corresponds to the continuum limit. (e) $lD_\perp$ vs $l$ and (f) $lD_\parallel$ vs $l$ are plotted as a function of $l$. Here, horizontal dashed line corresponds to $lD_\perp={\rm constant}$ and $lD_\parallel={\rm constant}$ for the case of full hydrodynamic screening among the beads belonging to a rod.} 
\label{fig: D_T}
\end{figure*}

The translational diffusion coefficients as a function of rod length $l$ are shown in figure \ref{fig: D_T} for different values of $\delta$. 
A closer look reveals that the behaviour of $D_\perp$ is closer to the curve with no hydrodynamic interaction considered, see figure~\ref{fig: D_T}(b), while there is minimal deviation of $D_\parallel$ from the curve with hydrodynamic interaction, figure~\ref{fig: D_T}(c).
Thus, the effect of hydrodynamic interaction among monomers is significant for the motion parallel to the rod axis, which is expected as the hydrodyanamic interaction among the monomers of a rod reduces the friction parallel to the rod axis\cite{wang2021diffusion}. 
Since $D_{\rm COM}$ is the combination of both of $D_\perp$ and $D_\parallel$ the dominant contribution comes from the parallel component for monomerically thin  rods\cite{brochard2000viscosity} and hence, the closeness of overall diffusion coefficient ($D_{\rm COM}$) to the reference level with hydrodynamic interaction considered. \

Since the rotational diffusion constant $D_{\rm R}$ is directly related to $D_{\perp}$\cite{wang2021diffusion}, for the case of rotation of rods perpendicular to its axis, the hydrodynamic interaction between the monomers of the same rod will be significantly reduced. 
As noted earlier, for rotational motion $\tau_{_R}/\tau_{_F} \approx 10^2-10^4$, while for the translational motion we find that $\tau_{_T}/\tau_{_F} \approx 10-10^2$, with $\tau_{_F}$ the relaxation time of the rod's translational motion. 
This result is in contrast to the study by Tsay \textit{et al.}\cite{tsay2006rotational}, which predicted a faster rotational relaxation time in comparison to the corresponding translational relaxation time. 
This may be  due to the fact that the nanorods used in fluorescence correlation spectroscopy (FCS) are coated with functional group like phytochelatin-related peptides which may have hindered the motion along the rod axis. \
The anisotropic constant, $k = D_\parallel / D_\perp$, is also calculated and plotted as a function of rod length, see figure~\ref{fig: D_T}(d). As expected, the deviation from $k=2$ gets stronger with the increase of rod length, also the deviation is larger for rods in more size-polydisperse fluid systems. Note that the value of $k<2$ and approaches 1 for rod length $l=2b$ (resembles dumbel) where $l$ matches the mean particle size of the fluid. The value of $k=1$ is expected for isotropic tracer like spherical nano-particles. 

%\section{Results and discussion}
%\label{sec: Results and discussion}

%%%%%%%%%%%%%%%%%%%%%%%%%%%%%
%%%%%%%%%%%%%%%%%%%%%%%%%%%%%
\section{Effect of density}\label{sec: density effect}
\begin{figure*}[htbp]
\includegraphics[width = 0.9\linewidth]{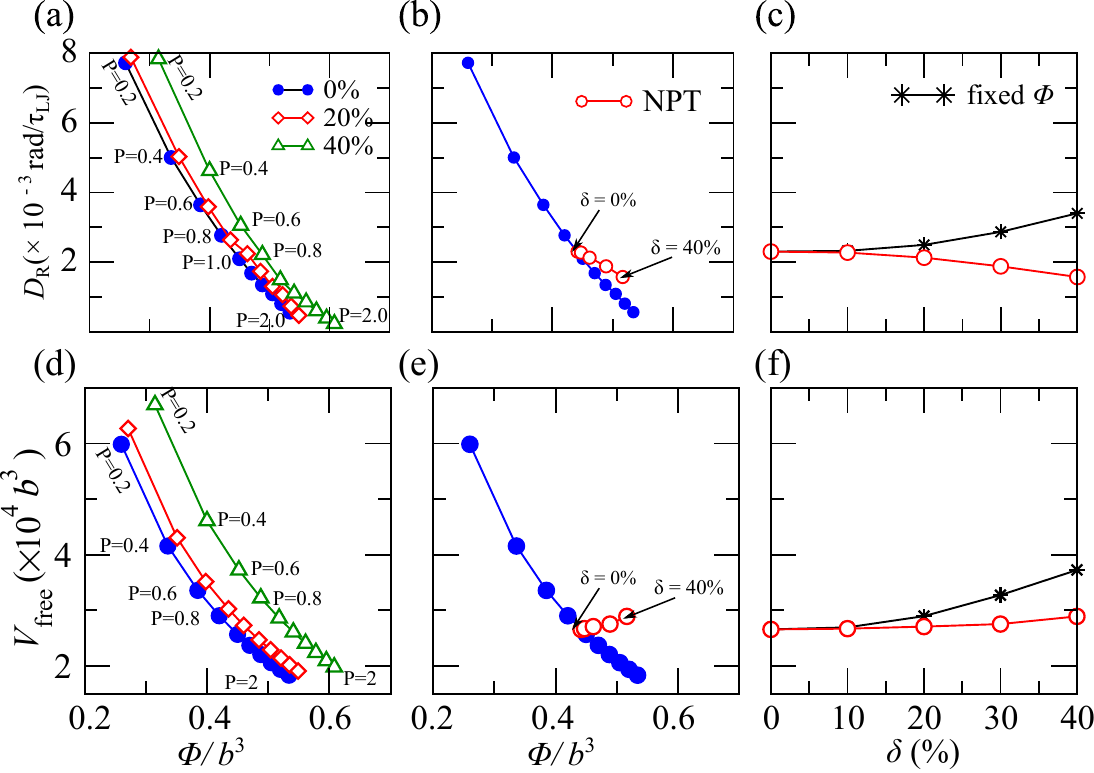}
\caption{(a) $D_{\rm R}$ vs $\phi$ for the rod in monodisperse and size-polydisperse HS fluids shown for $\delta=20\%, 40\%$. Under constant volume fraction diffusivity increases with increasing size-polydispersity. (b) Comparison of $D_{\rm R}$ vs $\phi$ for the rod ($l=12b$) in reference monodisperse HS fluid under constant volume fraction and the rod in size-polydisperse fluid whose $\phi$ changes with changing $\delta$ under constant pressure (NPT in the figure). (c) Comparion of $D_R$ vs $\phi$ for the rod ($l=12b$) embedded in two sets of size-polydisperse fluids: one under fixed $\phi$ and other with constant pressure. The error in the determination of $D_R$ values are at most $\pm 10^{-5}$. (d)-(f) Free volume, $V_{\rm{free}}$ vs $\phi$, corresponding to the systems in figure (a)-(c), respectively.}
\label{fig: fixed phi}
%\label{fig: phi_DR}
\end{figure*}

The change in size-polydispersity changes the volume fraction ($\phi$) of the system if the pressure and the temperature of the system remains unchanged (i.e., NPT conditions). 
In order to decouple the effect of changing $\phi$ from that of $\delta$, we compare the rotational dynamics of the rods embedded in  size-polydisperse HS fluids with those in monodisperse counterparts at the same/comparable value of $\phi$ by considering rod of length $l=12b$ (longest in this study). As detailed in section~\ref{sec: model-description}, for system with constant volume fraction, irrespective of the $\delta$ value the value of $\phi$ is held constant by adjusting volume. 
In figure~\ref{fig: fixed phi}(a), we plot $D_R$ as a function of $\phi$ for two different values of $\delta$ (20\% and 40\%) along with the reference monodisperse system ($\delta=0\%$). 
The rods in size-polydisperse fluids are more duffusive compared to the rods in monodisperse counterpart of comparable volume fractions and $D_R$ is larger for larger $\delta$. \

Again, in figure~\ref{fig: fixed phi}(b), we compare $D_R$ as a function of $\phi$ for the rods in monodisperse HS fluid ($\delta=0$) and that in size-polydisperse fluid prepared under constant pressure ($P=1$) where the volume fraction naturally varies in the range $0.44-0.53$ approximately as we vary $\delta$ from $0-40\%$. \
For the size-polydisperse system, as $\delta$ increases, we observed increased positive deviation of $D_R$ from the monodisperse counterpart, although the tracer diffusivity decreases with increasing $\delta$. It may be due to the inherent increase of volume fraction under constant pressure. In order to confirm this, we prepared a set of size-polydisperse systems with fixed $\phi$ (equal to that of reference monodisperse system) and fixed pressure (NPT in the figure) and compare their $D_{\rm R}$ as a function of $\delta$, see figure \ref{fig: fixed phi}-(c). 

Note that under the condition of fixed $\phi$, increasing size-polydispersity leads to lower pressure (or larger system volume to maintain constant $\phi$) than the monodisperse counterpart and thus the viscosity decreases~\cite{bird2002transport}, thereby increasing the value of $D_R$ in more polydisperse liquids. At the same time, we also observe that system with higher $\delta$ value have more absolute free volume, $V_{\rm{free}}=(1-\phi)\times \rm{system~ volume}$, compared to those with lower $\delta$ value for comparable values of $\phi$ under NPT conditions. In figure~\ref{fig: fixed phi} (d)-(f) the change in $V_{\rm free}$ correponding to figure~\ref{fig: fixed phi} (a)-(c), respectively, are shown. 
 
Comparasion of figure~\ref{fig: fixed phi}-(a) and \ref{fig: fixed phi}-(d) reveal that, at comparable $\phi$, higher polydispersity implies greater free volume under constant pressure. In figure \ref{fig: fixed phi}-(b) and \ref{fig: fixed phi}-(e), the NPT curve reveals, that increasing $\delta$ under fixed pressure ($P=1$) increases $\phi$ as well as $V_{\rm{free}}$. Again comparison of figure~\ref{fig: fixed phi}-(c) and (f) shows that in both size-polydisperse systems, i.e.~fixed $\phi$ and fixed pressure, $V_{\rm{free}}$ increases with increasing $\delta$, but it is relatively large for system with fixed $\phi$. And consequently a larger diffusivity with increasing $\delta$ for tracers in size-polydisperse HS fluid under constant volume fraction. Similar behaviour is observed for the translation motion also. 

%%%%%%%%%%%%%%%%%%%%%%%%% 
%%%%%%%%%%%%%%%%%%%%%%%%% 

\section{Conclusion} \label{sec: conclusion}

We have studied the dynamics of stiff rods embedded in  size-polydisperse HS fluids, focussing on the effect of degree of size disparity of the fluid particles (characterized by the polydispersity index, $\delta$) on the dynamics of the rods. The existing power law depedence of rotational and translational diffusion coefficients on the rod length as well as hydrodynamic interactions are also investigated. \\

Rotational dynamics slows down with increasing rod length in both size-polydisperse and reference monodisperse HS fluids. For the monodisperse case, the rotational diffusion constant, $D_R \propto l^{-\alpha}$, with $\alpha=3$ in agreement with the theoretical predictions. Also, the effect of hydrodynamic interactions among the beads of the rods are observed to be minimal in the case of rotation. Interestingly, for size-polydisperse HS fluid, $D_R \propto l^{-\alpha}$, where the exponent $\alpha$ varies in the range $3.0 - 3.2$ when the polydispersity index $\delta$ varies from $0-40\%$ (for system under constant NPT condition). 
We have ensured the accuracy of the simulated data by taking average of at least 15 replicas for each system; however, a much wider range of rod length may have to be considered to verify this power law dependence. 
It is evident from this study that the exponent $\alpha$ increases upon increasing the size-polydispersity of the fluid medium indicating slowing down of rod dynamics. It is due to the fact that with increasing size-polydispersity the system's volume fraction also increases and hence the drag force on the rods. \\
Furthermore, translational motion also slows down with both the increase of rod length and $\delta$, and the translational diffusion coefficient $D_{\rm COM} \propto l^{-\alpha}$, with $\alpha < 1$. The value of $\alpha=1$ represents the full hydrodynamic screening between the beads of the stiff-rod. The observed value of $\alpha<1$ indicates partial screening of the hydrodynamic interaction between the beads of the rod which is in contrast to that observed in the rotational motion. \\
Since the volume fraction of the system inherently changes with changing size-polydispersity under constant pressure, we isolated the effect of varying $\delta$  by preparing samples at fixed volume fraction of the system. This study further reveals that under both constant volume fraction and constant pressure conditions the system's absolute free volume increases with increasing $\delta$. Interestingly, with increasing $\delta$, the rotational dynamics slows down for system under constant pressure, whereas a faster dynamics is observed for the system with fixed $\phi$. Similar trend is also observed for the translational motion also.

In conclusion, increasing size-polydispersity inhibits tracer diffusion if the pressure of the system is held constant, while it enhances if the volume fraction is fixed. 
The current study provides a way to test the existing conflicting theories on dynamics of rod-like polymers in dynamically evolving crowded fluid-like medium, also it may have implications on the design of nano-rods with desired diffusivity. 

%%%%%%%%%%%%%%%%%%%%%%%%%%%%%%%%%%%%%%%%%%%%%%%%%%%%%%%%
\begin{center}
	\textbf{AUTHOR INFORMATION}
\end{center}
\textbf{Corresponding Author} \par
\noindent Lenin S.~Shagolsem -- {\it Department of Physics, National Institute of Technology Manipur, Imphal - 795004, India}; orcid.org/0000-0003-0836-7491; Email: lenin.shagolsem@nitmanipur.ac.in \\
\textbf{Authors} \par 
\noindent Thokchom Premkumar Meitei -- {\it Department of Physics, National Institute of Technology Manipur, Imphal - 795004, India} 

\medskip
\textbf{Notes} \par
The authors declare no competing financial interest.

%\medskip
%\textbf{Author's contribution} \par
%Both author conceptualized the work. JP performed simulations and drafted the manuscript. The corresponding author, LS edited the draft and ensure that all authors agree to the same.

\medskip
\textbf{Data Availability} \par
The data that support the findings of this study are available from the corresponding author upon reasonable request.
%%%%%%%%%%%%%%%%%%%%%%%%%%%%%%%%%%%%%%%%%%%%%%%%%%%%

%%%%%%%%%%%%%%%%%%%%%%%%%%%%%%%%%%%%%%%%%%%%%%%%%%%%%%%%
\bibliographystyle{aip}
\bibliography{ref}

\end{document}